\documentclass[aps,prl,twocolumn,superscriptaddress,longbibliography]{revtex4}
\pdfoutput=1
\usepackage{graphicx}
\usepackage{amssymb,amsmath}
\usepackage{bm}
\usepackage{dcolumn}
\usepackage{subfigure}
\usepackage{float}
\usepackage[OT1]{fontenc} 
\usepackage{url}
\usepackage{slashed}
\usepackage{color}
\usepackage{verbatim}
\usepackage{txfonts}
\usepackage{times}
\usepackage{graphicx}
\usepackage{hyperref}

\begin{document}
	
\title{Scrambling Ability of Quantum Neural Networks Architectures}

\author{Yadong Wu}
\affiliation{Institute for Advanced Study, Tsinghua University, Beijing, 100084, China}

\author{Pengfei Zhang}
\email{pengfeizhang.physics@gmail.com}
\affiliation{Institute for Quantum Information and Matter, California Institute of Technology, Pasadena, California 91125, USA}
\affiliation{Walter Burke Institute for Theoretical Physics, California Institute of Technology, Pasadena, California 91125, USA}

\author{Hui Zhai}
\email{hzhai@tsinghua.edu.cn}
\affiliation{Institute for Advanced Study, Tsinghua University, Beijing, 100084, China}

\begin{abstract}

In this letter we propose a general principle for how to build up a quantum neural network with high learning efficiency. Our stratagem is based on the equivalence between extracting information from input state to readout qubit and scrambling information from the readout qubit to input qubits. We characterize the quantum information scrambling by operator size growth, and by Haar random averaging over operator sizes, we propose an averaged operator size to describe the information scrambling ability for a given quantum neural network architectures, and argue this quantity is positively correlated with the learning efficiency of this architecture. As examples, we compute the averaged operator size for several different architectures, and we also consider two typical learning tasks, which are a regression task of a quantum problem and a classification task on classical images, respectively. In both cases, we find that, for the architecture with a larger averaged operator size, the loss function decreases faster or the prediction accuracy in the testing dataset increases faster as the training epoch increases, which means higher learning efficiency. Our results can be generalized to more complicated quantum versions of machine learning algorithms.        

\end{abstract}
	
\maketitle

Classical neural networks can extract information from the input, usually a high-dimensional vector, and encode the information into a number or a low-dimensional vector as output. Classical neural networks have found broad applications in both technology developments and scientific researches. For these applications, there are studies on how to design properly the architectures of neural networks, such as the number of layers, the number of neurons in each layer, and the activation functions, such that extracting information can be made most efficiently~\cite{classical_structure}. Quantum neural networks (QNN) also extract information from the input, usually a high-dimensional quantum wave function, and encode the information into one or a few read-out qubits. QNNs are considered one of the most promising applications in the near-term noise intermediate-scale quantum technology \cite{Preskill} and have attracted considerable attention recently \cite{QNN0,QNN1,QNN2,QNN3,QNN4,QNN5,QNN6,QNN7,QNN8,QNN9,QNN10,QNN11,QNN12,QNN13,QNN14,QNN15} . QNNs are made of local unitary quantum gates, and in practices, we should face the same problem of how we design the architectures of QNN properly. 

\begin{figure}[t]
	\includegraphics[width=.85\columnwidth]{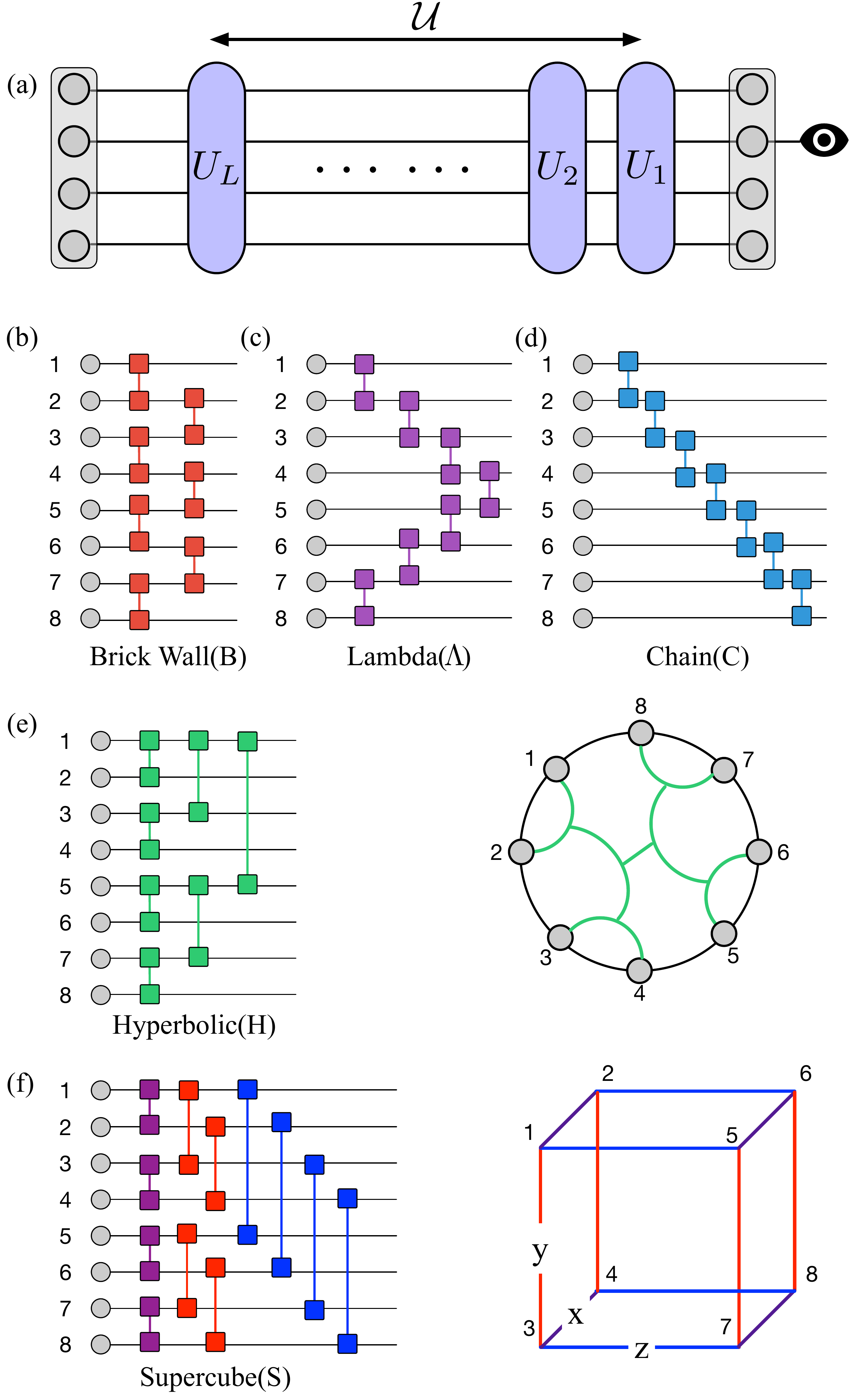}
	\caption{(a) The global structure of the QNN. Each $\hat{U}_l$ is called a unit in this work, which is chosen among one of the building blocks shown in (b-f). (b-f) Various typical building blocks for constructing the QNN, which are respectively called the ``brick wall"(B), ``Lambda" ($\Lambda$), ``Chain"(C), ``Hyperbolic"(H) and ``Supercube"(S) in this work.  } 
	\label{circuit}
\end{figure}

To be concrete, we consider QNN as shown in Fig.~\ref{circuit}(a). The dataset is denoted by $\{(|\psi^d\rangle,y^d\}$ ($d$ labels data), where $|\psi^l\rangle$ is a quantum wave function and $y^l$ is its label. The quantum circuit denoted by a unitary transformation $\hat{\mathcal{U}}$ is made of several local (say, two-qubit) quantum gates. There are various ways to construct $\hat{\mathcal{U}}$ with two-qubit gates, and different constructions correspond to different architectures. In the end, one measures the readout qubit-$r$, say, by measuring $\hat{\sigma}_x^r$, one can introduce the measurement operator $\hat{M}$ as  
\begin{equation}
\hat{M}=\hat{\sigma}^1_0\otimes\dots\otimes\hat{\sigma}^r_x \otimes\dots\hat{\sigma}^N_0, \label{measurement} 
\end{equation}
where the superscrip $i=1,\dots, N$ labels the qubits. Aside from the readout qubit $r$, no measurement is performed at other qubits, which are described by the identity matrix denoted by $\sigma^i_0$. The measurement yields a readout 
\begin{equation}
\tilde{y}^d=\langle \psi^d|\hat{\mathcal{U}}^\dag \hat{M} \hat{\mathcal{U}}|\psi^d\rangle. \label{readout}
\end{equation}
A loss function is designed to measure how close $\tilde{y}^d$ is to $y^d$, and one trains the parameters in the two-qubit gates to minimize the loss function. During training, the QNN can also make predictions on the test dataset. Therefore, for a given task and dataset, and by averaging over different initializations, the loss or the accuracy as a function of training epoch mostly depends on the architecture of QNN. The issue addressed in this work is whether there is a guiding principle for designing the most efficient architecture in learning, that is, as the training epoch increases, the decreasing of the loss or the increasing of the accuracy is the fastest.  

In the case of classical neural networks, there is always information loss from the input layer to the output layer. However, for QNN, the total information is conserved during unitary transformation through the quantum circuit. Note that a unitary transformation is reversible, when we say a QNN encodes the information from the input wave function to the readout qubit, it is equivalent to say that the QNN scrambles the information from the readout qubit to all input qubits. Thus, the efficiency of extracting information is equivalent to the efficiency of scrambling information. Lots of studies in the past few years have established several related quantities to characterize quantum information scrambling, such as the out-of-time-ordered correlator~\cite{OTOC1,OTOC2,OTOC3,OTOC4,OTOC5,OTOC6}, the tripartite information~\cite{I31,I32,I33,I34} and the operator size growth~\cite{OSG1,OSG2,OSG3,OSG4,OSG5,OSG6,OSG7,OSG8,OSG9,OSG10,OSG11}. Recently, the tripartite information has also been used to reveal universal features in the training dynamics of QNN~\cite{I3_QNN}. This work focuses on the architecture and the main results are two folds:

i) We propose a quantity based on the operator size to characterize the information scrambling ability of a QNN architecture. 

ii) We show that the scrambling ability quantified in this way is positively correlated with the learning efficiency of the QNN architecture. 

\textit{Architectures.} We demonstrate our results using several different architectures shown in Fig.~\ref{circuit}(b-f) as examples. The entire quantum circuit $\hat{\mathcal{U}}$ is made of a number of units, i.e. $\hat{\mathcal{U}}=\hat{U}_1\hat{U}_2\dots\hat{U}_L$, as shown in Fig.~\ref{circuit}(a). In each unit, $\hat{U}_l$ contains a number of two-qubit gates $\hat{u}_{ij}$, and it ensures that every qubit is operated at least once. $\hat{u}_{ij}$ denotes a two-qubit gate acting on qubit-$i$ and -$j$. Each $\hat{u}_{ij}$ is parameterized as 
\begin{equation}
\hat{u}_{ij}=e^{\sum_{k}\alpha^k_{ij}\hat{g}_k},
\end{equation}     
where $\hat{g}_k$ are $SU(4)$ generators and $\alpha^k_{ij}$ are parameters. In a QNN, these parameters need to be determined by training. How to arrange these $\hat{u}_{ij}$ to form $\hat{U}_l$, and then to form $\hat{\mathcal{U}}$, is referred to as the architecture. 

Fig.~\ref{circuit}(b-f) show architectures considered in this work. For cases shown in Fig.~\ref{circuit}(b-d), all qubits are aligned along a one-dimensional line and all gates operator on two neighboring qubits. They differ by the ordering of these gates, and they are called Brick Wall (B), Lambda ($\Lambda$), and Chain (C) as what they look like. For the case shown in Fig.~\ref{circuit}(e), all qubits sit in a one-dimensional circle, and the way they interact is reminiscent of the hyperbolic geometry, for which it is called Hyperbolic (H). Finally, for the case shown in Fig.~\ref{circuit}(f), qubits sit at the corners of a three-dimensional cube. The two-qubit gates first act on four pairs of neighboring gates along $x$, and then four pairs of neighboring gates along $y$ and finally four pairs of neighboring gates along $z$. Below we explicitly show the scrambling ability and its correlation with learning ability using these architectures, however, we emphasize that we have tried more generic architectures and our conclusions below hold for general architectures.

\textit{Operator Size.} Now we briefly introduce the operator size~\cite{OSG1,OSG2,OSG3,OSG4,OSG5,OSG6,OSG7,OSG8,OSG9,OSG10,OSG11}. Let us consider a system with $N$-qubit and an operator $\hat{\mathcal{O}}$ in this system. Generally, we can expand the operator as
\begin{equation}
\hat{\mathcal{O}}=\sum_{{\bm \alpha}}c_{{\bm \alpha}}\hat{\sigma}^1_{\alpha_1}\otimes\hat{\sigma}^2_{\alpha_2}\dots\otimes\hat{\sigma}^N_{\alpha_N}, \label{expansion}
\end{equation}
where $\hat{\sigma}^i_{\alpha_i}$ with subscript $\alpha_i=0,1,2,3$ respectively denotes identity ($\alpha_i=0$) and three Pauli matrices $\hat{\sigma}_{x,y,z}$. Here ${\bm \alpha}$ denotes a set $\{\alpha_1,\alpha_2,\dots,\alpha_N\}$, and we use $l({\bm \alpha})$ to denote the number of non-zero elements in the set ${\bm \alpha}$, i.e. the number of operators in $\hat{\sigma}^1_{\alpha_1}\otimes\hat{\sigma}^2_{\alpha_2}\dots\otimes\hat{\sigma}^N_{\alpha_N}$ that are not identity. Then, the size of an operator is defined as
\begin{equation}
\text{Size}(\hat{\mathcal{O}})=\sum_{{\bm \alpha}}|c_{{\bm \alpha}}|^2 l({\bm \alpha}).
\end{equation}
In the most general case, there are totally $4^N$ terms in the expansion Eq.~\ref{expansion}. Here we give two examples. If we consider the measurement operator $\hat{M}$ defined in Eq.~\ref{measurement}, we have $\text{Size}(\hat{\mathcal{O}})=1$. If we consider a uniform distribution among all $4^N-1$ traceless operators, with $|c_{{\bm \alpha}}|^2=1/(4^N-1)$, then
\begin{equation}
\text{Size}(\hat{\mathcal{O}})=\frac{1}{4^N-1}\sum_{n=1}^{N}\frac{N!}{n!(N-n)!}3^nn\approx\frac{3N}{4}.
\end{equation}
Furthermore, if we consider the situation that, among $N$-qubit, operators on a fraction of $\alpha N$ qubits ($\alpha<1$) are uniformly distributed among $\hat{\sigma}_{0,\dots,4}$ and operators on the rest $(1-\alpha)N$ qubits are always identity. Then, the operator size is reduced to $3\alpha N/4$. 

We present an argument to bring out the connection between operator size and the learning ability of a QNN. Let us consider the operator $\hat{M}^\prime=\hat{\mathcal{U}}^\dag \hat{M} \hat{\mathcal{U}}$ in Eq. \ref{readout}. Initially, $\hat{M}$ operator is not identity only at the measurement qubit-$r$, however, because $\hat{\mathcal{U}}$ does not commute with $\hat{M}$, $\hat{M}^\prime$ can also be one of the three Pauli matrices on other qubits and the operator size increases. Generally, when the operator $\hat{\mathcal{U}}$ becomes more and more complicated as the depth of QNN increases, the operator size of $\hat{M}^\prime$ increases. However, if $\text{Size}(\hat{M}^\prime)$ is not sufficiently large, there is still a large probability that $\hat{M}^\prime$ takes identity operator on some qubits. Since $\hat{M}^\prime$ acts on the input state, and if the operator $\hat{M}^\prime$ is nearly identity on some qubits, the QNN can hardly extract information from the input wave function at those qubits. Therefore, a necessary condition for accurate learning is that $\text{Size}(\hat{M}^\prime)$ reaches a sufficient large value.

\begin{figure}[t]
	\includegraphics[width=.85\columnwidth]{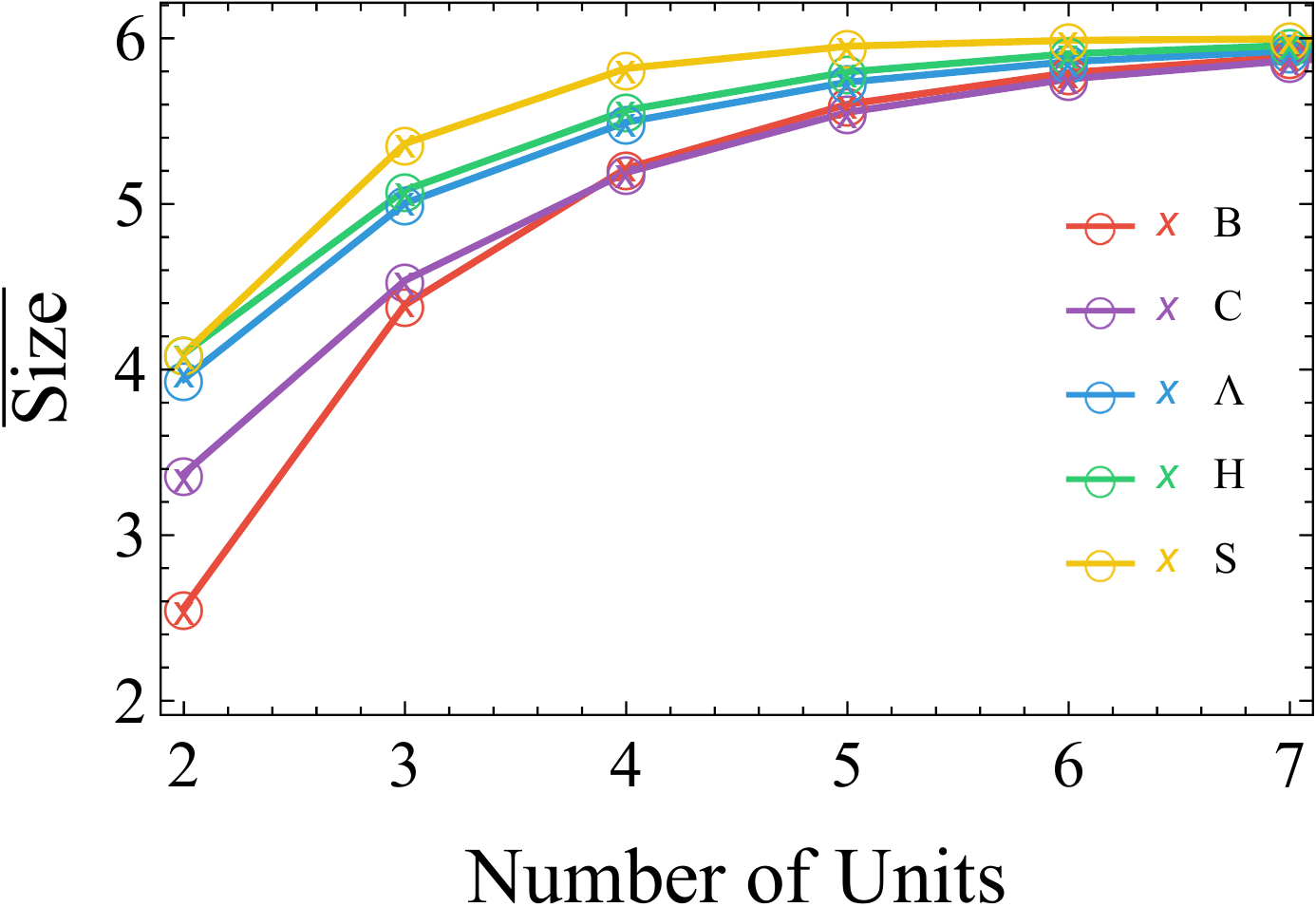}
	\caption{The Haar-random-averaged operator size $\overline{\text{Size}}$ defined in Eq.~\ref{scrambling_arc} for different architectures. For each architecture, all units share the same structure chosen as one of the cases shown in Fig.~\ref{circuit}(b-e) with (f) a little different \cite{Scube} and labeled by the same label introduced in Fig.~\ref{circuit}(b-f). The horizontal axis is the number of units. Cross markers with different labels are obtained by numerical simulations and the solid lines with empty circles are obtained with analytical formula. Here we have taken the number of qubits $N=8$. } 
	\label{size}
\end{figure} 

$\text{Size}(\hat{M}^\prime)$ depends on both the architecture and the parameters of the unitary $\hat{\mathcal{U}}$. Since for a QNN, the parameters keep updating during training but the architecture is fixed as a prior, we would like to have a quantity that only depends on the architecture. To this end, we propose to consider an averaged operator size  
\begin{equation}
\overline{\text{Size}}=\int d\hat{\mathcal{U}} \text{Size}(\hat{\mathcal{U}}^\dag\hat{M}\hat{\mathcal{U}}). \label{scrambling_arc}
\end{equation}
Here $\int d\hat{\mathcal{U}}$ means Haar random average overall two-qubit gates in $\hat{\mathcal{U}}$. Since the parameters in $\hat{\mathcal{U}}$ have been averaged over, $\overline{\text{Size}}$ defined by Eq.~\ref{scrambling_arc} only depends on the architecture. This quantifies characterizes that for generic parameters, how fast the operator size grows in a given QNN architecture. We propose to use this parameter to quantify the ability to scramble quantum information for a given architecture. We argue that for an architecture with larger $\overline{\text{Size}}$, it is easier to reach a suitable parameter such that $\text{Size}(\hat{\mathcal{U}}^\dag\hat{M}\hat{\mathcal{U}})$ is large enough that ensures efficient information extraction from the input wave functions.  

The Haar random average can also simplify the calculation of the operator size. For instance, let us consider a two-qubits system and an operator $\hat{\sigma}_x \otimes \hat{\sigma}_0$. Expanding $\hat{\mathcal{U}}^\dagger \hat{\sigma}_x \otimes \hat{\sigma}_0 \hat{\mathcal{U}}$ as Eq.~\ref{expansion}, and after averaging over the Haar random unitary, the weight $c_{\bm{\alpha}}$ (${\bm \alpha}=(\alpha_1,\alpha_2)$) reads~\cite{OSG1}
\begin{equation}\label{fact}
\overline{|c_{\bm{\alpha}}|^2}=\frac{1-\delta_{\alpha_10}\delta_{\alpha_20}}{15}.
\end{equation}
Consequently, the probability of having a non-identity operator only on the first or only on the second site is $1/5$, and the probability for having non-identity operators on both sites is $3/5$. Based on Eq.~\eqref{fact}, for any QNN with $\hat{\mathcal{U}}$ composited by two-qubits gates, the operator size growth can be explicitly deduced as the depth of the QNN increases.

We compute $\overline{\text{Size}}$ defined in Eq.~\ref{scrambling_arc} for different architectures shown in Fig.~\ref{circuit} and the results are shown in Fig.~\ref{size}. The results show the ordering of $\overline{\text{Size}}$ as $\text{(S)}>\text{(H)}\approx (\Lambda) >\text{(C)}\approx\text{(B)}$. Especially, it is clear that the supercube (S) performs obviously better than others. And the difference between different architectures is the most significant for intermediate QNN depth. When the number of units is too small (e.g. $\sim 2$) and the QNN is too shallow, the unitary is not complicated enough that a local operator cannot be sufficiently scrambled for all architectures. On the other hand, when the number of units is large enough (e.g. $\sim 7$) and the QNN is deep enough, the unitary is sufficiently complicated for all architectures and $\overline{\text{Size}}$ for all cases approach $3N/4$ ($=6$ for $N=8$ considered here), and their differences also become insignificant.

\begin{figure}[t]
	\includegraphics[width=\columnwidth]{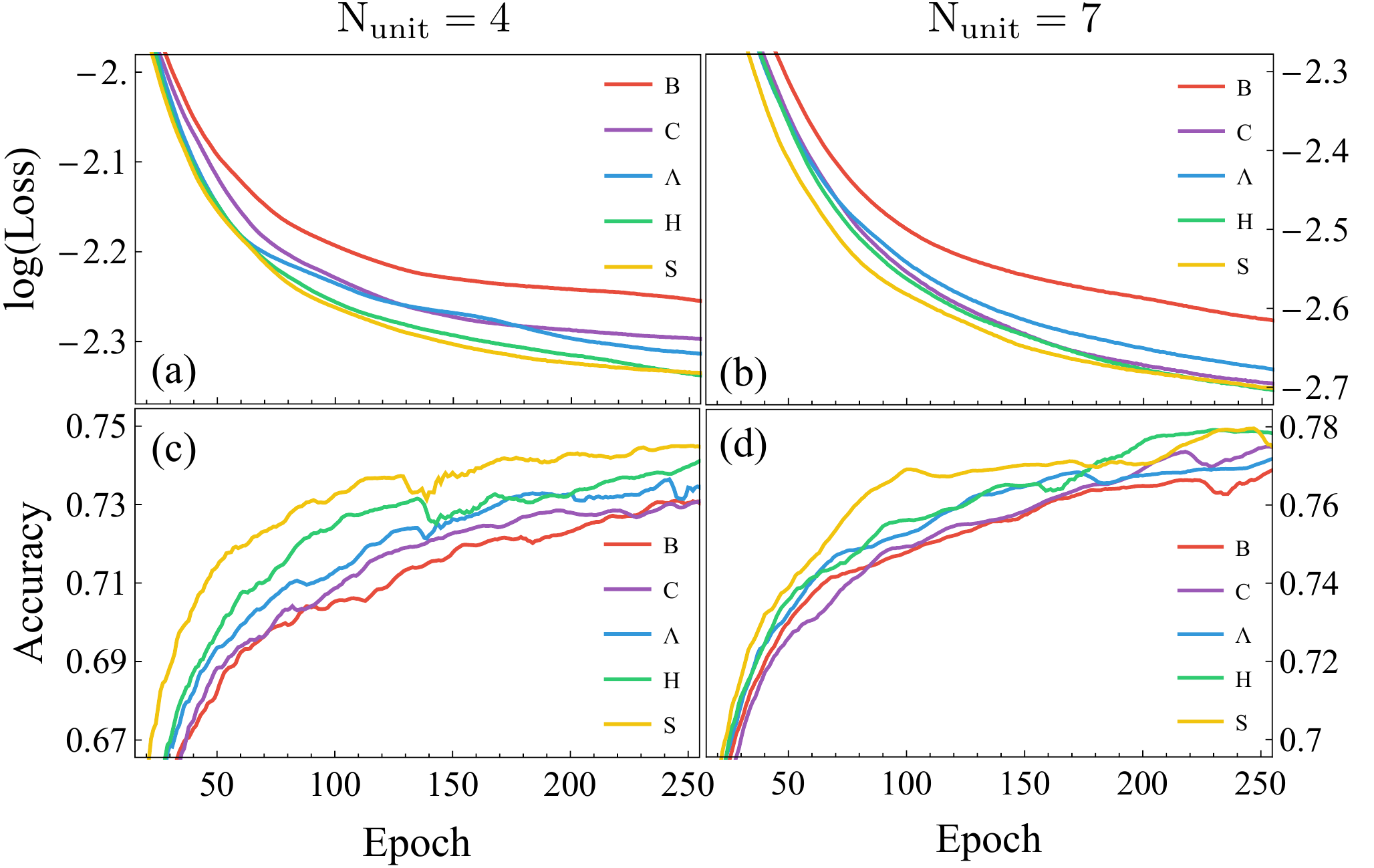}
	\caption{Performance of different architectures for two different tasks. (a-b) Loss as a function of training epoch for the information recovering task on the quantum spin problem. (c-d) Prediction accuracy as a function of training epoch for the second classification problem of RGB images of numbers. The number of units $N_{\text{unit}}=4$ for (a) and (c) and $N_{\text{unit}}=7$ for (b) and (d). In both cases, the results have been averaged over $10$ different initializations. } 
	\label{examples}
\end{figure}

\textit{Learning Efficiency.} To relate the learning efficiency to the scrambling ability defined above, we consider two typical training tasks. The first is a regression task of information recovering in a quantum system. Let us consider an unknown initial product state $|\phi^d\rangle$, its total magnetization is given by
\begin{equation}
M_z^d=\frac{1}{N}\langle \phi^d|\sum\limits_{i=1}^{N}\hat{\sigma}^i_z|\phi^d\rangle.
\end{equation}
Now let us consider a chaotic Hamiltonian 
\begin{equation}
\hat{H}=\sum\limits_{\langle ij\rangle}\sum\limits_{\alpha=x,y,z}J^{ij}_\alpha\hat{\sigma}^i_\alpha\hat{\sigma}^j_\alpha+\sum\limits_{i}\sum\limits_{\alpha=x,y,z}h^i_\alpha\hat{\sigma}^i_\alpha,
\end{equation}
where $J^{ij}_\alpha$ and $h^i_\alpha$ are a set of randomly chosen parameters. We evolve $|\phi^d\rangle$ with this Hamiltonian for sufficient long time to ensure a chaotic unitary dynamics, which yields $|\psi^d\rangle=e^{-i\hat{H}t}|\phi^d\rangle$. For QNN, the training dataset is taken as $\{\{|\psi^d\rangle,y^d\},d=1,\dots,N_\text{D}\}$, where $y^d$ is taken as the total magnetization $y^d=M^d_z$, and $N_\text{D}$ is the number of dataset. The loss function is taken as
\begin{equation}
\mathcal{L}=\frac{1}{N_\text{D}}\sum\limits_{d=1}^{N_\text{D}}|\tilde{y}^d-y^d|,\\
\end{equation}
where $\tilde{y}^d$ is the readout of QNN given by Eq.~\ref{readout} with input $|\psi^d\rangle$. In Fig.~\ref{examples}(a-b) we show how the loss function decreases as the training epoch increases. The trained QNN supposedly can recover the magnetization information of the initial state from the final state after a chaotic evolution. 

The second task is a classification task of recognizing classical images. We take large numbers of RGB images with either a number $6$ or a number $9$ embedded in the background. Each image contains $16\times 16=256$ pixels. Considering a system with $N=8$ qubits, there are totally $2^8=256$ bases in the Hilbert space. A general wave function can be expanded in terms of these $256$ bases. Each pixel corresponds to a base, and the information of each pixel is encoded into the coefficient of its corresponding base~\cite{RGB}. In this way, for each image, we generate a wave function $|\psi^d\rangle$ as input. The label is taken as $y^d=0$ if the image contains the number $6$ and $y^d=1$ if the image contains the number $9$. The readout of the QNN $\tilde{y}^d$ is also given by Eq.~\ref{readout} with the input $|\psi^d\rangle$. In this case, the loss function is taken as the cross-entropy between $y^d$ and $p^d$, and since $\tilde{y}^d$ lies between $[-1,1]$, we define $p^d$ as $(1+\tilde{y}^d)/2$ such that it lies in the range of $[0,1]$. Then the loss function is given by 
\begin{equation}
\mathcal{L}=-\frac{1}{N_\text{D}}\sum\limits_{d=1}^{N_\text{D}}[-y^d\log p^d-(1-y^d)\log(1-p^d)].
\end{equation}
After leaning, we let the QNN to make predictions on a testing dataset $\{\{|\psi^d\rangle,y^d\},d=1,\dots,N_\text{test}\}$. For each input $|\psi^d\rangle$, a trained QNN returns a prediction $\tilde{y}^d$ given by Eq.~\ref{readout}. Now we interpret the prediction as the number $9$ with $p^d=1$ if $\tilde{y}^d>0$, and as the number $6$ with $p^d=0$ if $\tilde{y}^d<0$. Then, we can obtain an accuracy as
\begin{equation}
\frac{1}{N_\text{D}}\sum\limits_{d=1}^{N_\text{D}}|p^d-y^d|.
\end{equation} 
In Fig.~\ref{examples}(c-d) we also show how the accuracy increases as the training epoch increases. 

The results shown in Fig.~\ref{examples} have been averaged over a few runs with different initializations, and therefore, their differences mainly reflect the differences in learning efficiency between different architectures. In Fig.~\ref{examples}(a-b), we show that in the first task, for most training epochs, the loss function is ordered as $(S)< (H)\lesssim (\Lambda) \lesssim (C)<(B)$. In Fig.~\ref{examples}(c-d), we show that in the second task, for most training epochs, the accuracy is ordered as $(S)>(H)\gtrsim (\Lambda)>(C)\gtrsim (B)$. Both orders are consistent with the order of $\overline{\text{Size}}$ defined for different architectures. This means that for a fixed target loss value or prediction accuracy, the architecture with the largest $\overline{\text{Size}}$ can reach this target with the smallest training epoch. In this sense, we consider this architecture as the most efficient one.  Therefore, these examples support our intuition of the positive correlation between scrambling ability and learning efficiency.

We also note that this correlation is most pronounced for intermediate training epochs and for intermediate depths of the QNN. This is because $\overline{\text{Size}}$ quantifies the scrambling ability of architectures with generic parameters, but for sufficiently long training, the QNN can always reach the optimal parameters. Also, for sufficiently deep QNN, all architectures with generic parameters can always lead to the most scrambled operators, whose size reaches the saturation value, as one can see from Fig.~\ref{size}. Therefore, the differences in learning efficiency also become less significant, as one can see by comparing Fig.~\ref{examples}(b)(d) with (a)(c). 

\textit{Outlook.} To the best of our knowledge, this work is the first attempt to understand how to design the most efficient architectures in QNN. Our design principle is based on quantum information scrambling in a quantum circuit, described by the operator size growth. We propose a quantity to quantify the scrambling ability of a QNN architecture, which is based on how fast the size of a local operator grows under generic unitary transformations generated by the quantum circuit. We conjecture the positive correlation between this quantity and the learning ability of the QNN, and the conjecture is confirmed by two typical learning tasks. Our discussion is so far limited to the quantum version of fully connected neural networks, and in the future, it can be generalized to other quantum versions of neural networks, such as quantum convolutional neural networks \cite{QCNN1, QCNN2, QCNN3}, quantum recurrent neural networks \cite{QRNN1,QRNN2}, and quantum autoencoders \cite{QAE1,QAE2,QAE3}.     

\textit{Acknowledgment.} This work is supported by Beijing Outstanding Young Scientist Program, NSFC Grant No. 11734010, MOST under Grant No. 2016YFA0301600.


\begin{thebibliography}{99}
\bibitem{Scube}
For supercube circuits (S), we set the last unit $\hat{U}_1$ the same as the Hyperbolic unit Fig.~\ref{circuit}(e). Other units $\hat{U}_{l\geq 2}$ are choosen such that the circuit $\hat{U}_{L}\cdots \hat{U}_{2}$ repeats the structure of Fig.\ref{circuit}(f) and each unit contains $7$ two-qubit gates. Such a structure eliminates redundant gates in $U_1$ and is found to be optimal for the operator size growth.
\end{thebibliography}
\end{document}